\documentclass[journal]{IEEEtran}
\usepackage{color}
\usepackage{graphicx}
\usepackage{cite}
\usepackage{url}
\usepackage{subfig}
\usepackage{array}

\makeatletter

\newcommand{\noun}[1]{\textsc{#1}}
\providecommand{\tabularnewline}{\\}

\@ifundefined{showcaptionsetup}{}{%
 \PassOptionsToPackage{caption=false}{subfig}}
\makeatother

\begin{document}

\title{Mobile Big Data Analytics\\Using Deep Learning and Apache Spark}

\author{Mohammad Abu Alsheikh, Dusit Niyato, Shaowei Lin, Hwee-Pink Tan, and Zhu Han
\thanks{M.~Abu~Alsheikh is with the School of Computer Engineering, Nanyang Technological University, Singapore 639798, and also with the Sense and Sense-abilities Programme, Institute for Infocomm Research, Singapore 138632 (e-mail: stumyhaa@i2r.a-star.edu.sg). D. Niyato is with the School of Computer Engineering, Nanyang Technological University, Singapore 639798 (e-mail: dniyato@ntu.edu.sg). S. Lin is with the School of Engineering Systems and Design Pillar, Singapore University of Technology and Design, Singapore 487372 (e-mail: shaowei\_lin@sutd.edu.sg). H.-P. Tan is with the School of Information Systems, Singapore Management University, Singapore 188065 (e-mail: hptan@smu.edu.sg). Z. Han is with the School of Electrical and Computer Engineering, University of Houston, USA  77004 (e-mail: zhan2@uh.edu).}
}

\maketitle
\begin{abstract}
The proliferation of mobile devices, such as smartphones and Internet of Things (IoT) gadgets, results in the recent mobile big data (MBD) era. Collecting MBD is unprofitable unless suitable analytics and learning methods are utilized for extracting meaningful information and hidden patterns from data. This article presents an overview and brief tutorial of deep learning in MBD analytics and discusses a scalable learning framework over Apache Spark. Specifically, a distributed deep learning is executed as an iterative MapReduce computing on many Spark workers. Each Spark worker learns a partial deep model on a partition of the overall MBD, and a master deep model is then built by averaging the parameters of all partial models. This Spark-based framework speeds up the learning of deep models consisting of many hidden layers and millions of parameters. We use a context-aware activity recognition application with a real-world dataset containing millions of samples to validate our framework and assess its speedup effectiveness.
\end{abstract}


\section{Introduction}

Mobile devices have matured as a reliable and cheap platform for collecting data in pervasive and ubiquitous sensing systems. Specifically, mobile devices are (a)~sold in mass-market chains, (b)~connected to daily human activities, and (c)~supported with embedded communication and sensing modules. According to the latest traffic forecast report by Cisco Systems \cite{forecast2015cisco}, half a billion mobile devices were globally sold in 2015, and the mobile data traffic grew by 74\% generating 3.7 exabytes (1~exabyte = $10^{18}$~bytes) of mobile data per month. \emph{Mobile big data} (MBD) is a concept that describes a massive amount of mobile data which cannot be processed using a single machine. MBD contains useful information for solving many problems such as fraud detection, marketing and targeted advertising, context-aware computing, and healthcare. Therefore, MBD analytics is currently a high-focus topic aiming at extracting meaningful information and patterns from raw mobile data.

Deep learning is a solid tool in MBD analytics. Specifically, deep learning (a)~provides high-accuracy results in MBD analytics, (b)~avoids the expensive design of handcrafted features, and (c)~utilizes the massive unlabeled mobile data for unsupervised feature extraction. Due to the curse of dimensionality and size of MBD, learning deep models in MBD analytics is slow and takes anywhere from a few hours to several days when performed on conventional computing systems. Arguably, most mobile systems are not delay tolerant and decisions should be made as fast as possible to attain high user satisfaction. 

To cope with the increased demand on scalable and adaptive mobile systems, this article presents a tutorial on developing a framework that enables time-efficient MBD analytics using deep models with millions of modeling parameters. Our framework is built over Apache Spark~\cite{spark2015apache} which provides an open source cluster computing platform. This enables distributed learning using many computing cores on a cluster where the continuously accessed data is cached to running memory, thus speeding up the learning of deep models by several folds. To prove the viability of the proposed framework, we implement a context-aware activity recognition system~\cite{lara2013survey} on a computing cluster and train deep learning models using millions of data samples collected by mobile crowdsensing. In this test case, a client request includes accelerometer signals and the server is programmed to extract the underlying human activity using deep activity recognition models. We show significant accuracy improvement of deep learning over conventional machine learning methods, improving $9\%$ over random forests and $17.8\%$ over multilayer perceptions from~\cite{weiss2012impact}. Moreover, the learning time of deep models is decreased as a result of the paralleled Spark-based implementation compared to a single machine computation. For example, utilizing $6$ Spark workers can speedup the learning of a $5$-layer deep model of 20 million parameters by $4$~folds as compared to a single machine computing.

The rest of this article is organized as follows. Section~\ref{sec:MBD} presents an overview of MBD and discusses the challenges of MBD analytics. Section~\ref{sec:deep_learning} discusses the advantages and challenges of deep learning in MBD analytics. Then, Section~\ref{sec:DL_MBD_spark} proposes a Spark-based framework for learning deep models for time-efficient MBD analytics within large-scale mobile systems. Section~\ref{sec:case_study} presents experimental analysis using a real-world dataset. Interesting research directions are discussed in Section~\ref{sec:future_work}. Finally, Section~\ref{sec:conclusions} concludes the article.

\section{Mobile Big Data (MBD): Concepts and Features}\label{sec:MBD}

This section first introduces an overview of MBD and then discusses the key characteristics which make MBD analytics challenging.

\subsection{The Era of MBD}

\begin{figure*}
\begin{centering}
\includegraphics[width=0.75\paperwidth,trim=1cm 1cm 1cm 0cm]{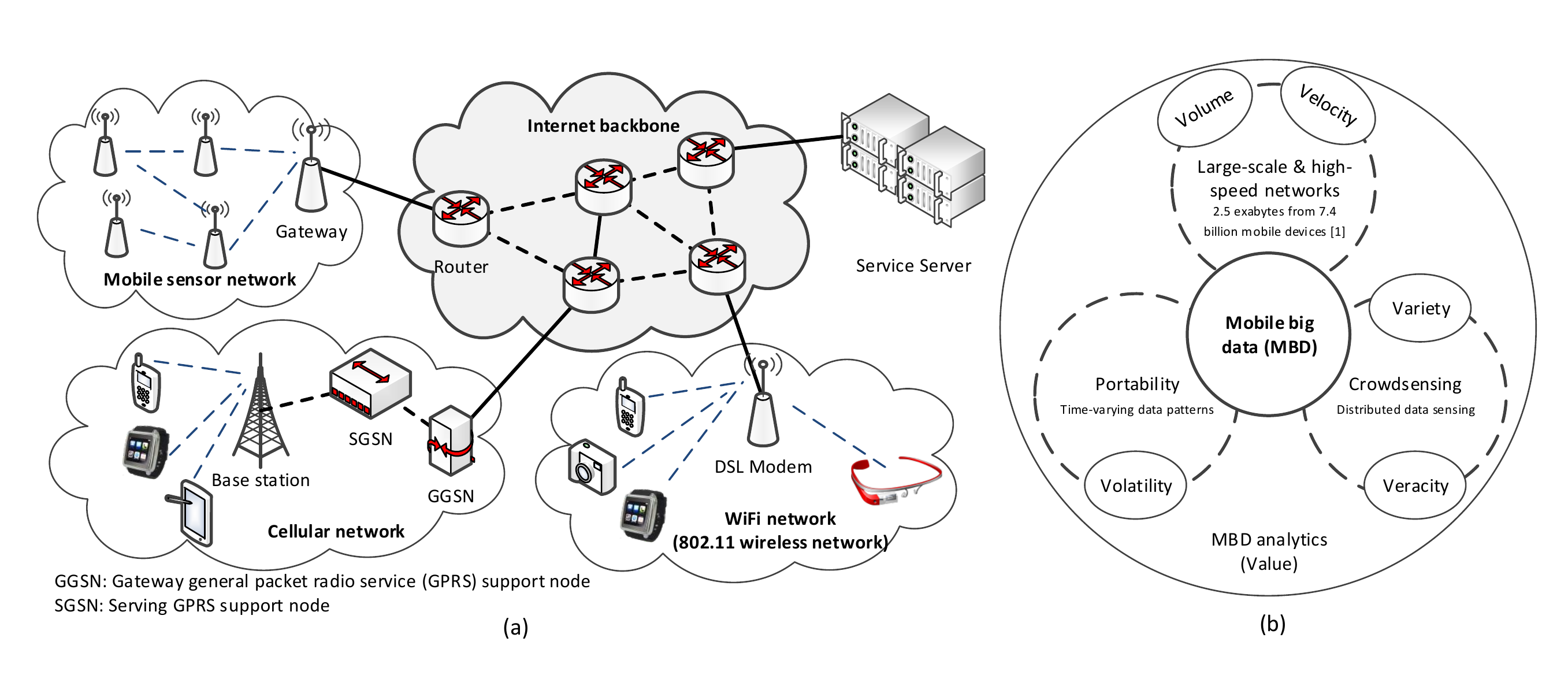}
\par\end{centering}

\caption{Illustration of the MBD era. (a)~Typical architecture of a modern mobile network connecting smartphones, wearable computers, and IoT gadgets. (b)~Main technological advances behind the MBD era.\label{fig:system_architecture}}
\end{figure*}

Figure~\ref{fig:system_architecture}~(a) shows a typical architecture of large-scale mobile systems used to connect various types of portable devices such as smartphones, wearable computers, and IoT gadgets. The widespread installation of various types of sensors, such as accelerometer, gyroscope, compass, and GPS sensors, in modern mobile devices allows many new applications. Essentially, each mobile device encapsulates its service request and own sensory data in stateless data-interchange structure, e.g., Javascript object notation (JSON) format. The stateless format is important as mobile devices operate on different mobile operating systems, e.g., Android, IOS, and Tizen. Based on the collected MBD, a service server utilizes MBD analytics to discover hidden patterns and information. The importance of MBD analytics stems from its roles in building complex mobile systems that could not be assembled and configured on small datasets. For example, an activity recognition application~\cite{lara2013survey,perera2014context} uses embedded accelerometers of mobile devices to collect proper acceleration data about daily human activities. After receiving a request, the service server maps the accelerometer data to the most probable human activities which are used to support many interactive services, e.g., healthcare, smart building, and pervasive games.

MBD analytics is more versatile than conventional big data problems as data sources are portable and data traffic is crowdsourced. MBD analytics deals with massive amount of data which is collected by millions of mobile devices. Next, we discuss the main characteristics of MBD which complicate data analytics and learning on MBD compared to small datasets.

\subsection{Challenges of MBD Analytics\label{sub:MBD_characteristics}}

Figure~\ref{fig:system_architecture}~(b) shows the main recent technologies that have produced the challenging MBD era: large-scale and high-speed mobile networks, portability, and crowdsourcing. Each technology contributes in forming the MBD characteristics
in the following way.
\begin{itemize}
\item Large-scale and high-speed mobile networks: The growth of mobile devices and high-speed mobile networks, e.g., WiFi and cellular networks, introduces massive and contentiously-increasing mobile data traffic. This has been reflected in the following MBD aspects:

\begin{itemize}
\item \emph{MBD is massive (volume)}. In 2015, 3.7 exabytes of mobile data was generated per month which is expected to increase through the coming years \cite{forecast2015cisco}.
\item \emph{MBD is generated at increasing rates (velocity)}. MBD flows at a high rate which impacts the latency in serving mobile users. Long queuing time of requests results in less satisfied users and increased cost of late decisions.
\end{itemize}
\item Portability: Each mobile device is free to move independently among many locations. Therefore, \emph{MBD is non-stationary (volatility)}.
Due to portability, the time duration for which the collected data is valid for decision making can be relatively short. MBD analytics should be frequently executed to cope with the newly collected data samples.
\item Crowdsourcing: A remarkable trend of mobile applications is crowdsourcing for pervasive sensing which includes a massive data collection from many participating users. Crowdsensing differs from conventional mobile sensing systems as the sensing devices are not owned by one institution but instead by many individuals from different places. This has introduced the following MBD challenges:

\begin{itemize}
\item \emph{MBD quality is not guaranteed (veracity)}. This aspect is critical for assessing the quality uncertainty of MBD as mobile systems do not directly manage the sensing process of mobile devices. Since most mobile data is crowdsourced, MBD can contain low quality and missing data samples due to noise, malfunctioning or uncalibrated sensors of mobile devices, and even intruders, e.g., badly-labeled crowdsourced data. These low quality data points affect the analytical accuracy of MBD.
\item \emph{MBD is heterogeneous (variety)}. The variety of MBD arises because the data traffic comes from many spatially distributed data sources, i.e., mobile devices. Besides, MBD comes in different data types due to the many sensors that mobile devices support. For example, a triaxial accelerometer generates proper acceleration measurements while a light sensor generates illumination values.
\end{itemize}
\end{itemize}
MBD analytics (\emph{value}) is mainly about extracting knowledge and patterns from MBD. In this way, MBD can be utilized for providing better service to mobile users and creating revenues for mobile businesses. The next section discusses deep learning as a solid tool in MBD analytics.

\section{Deep Learning in MBD Analytics}\label{sec:deep_learning}

\begin{figure}
\begin{centering}
\includegraphics[width=0.95\columnwidth,trim=1cm 0.8cm 1cm 0cm]{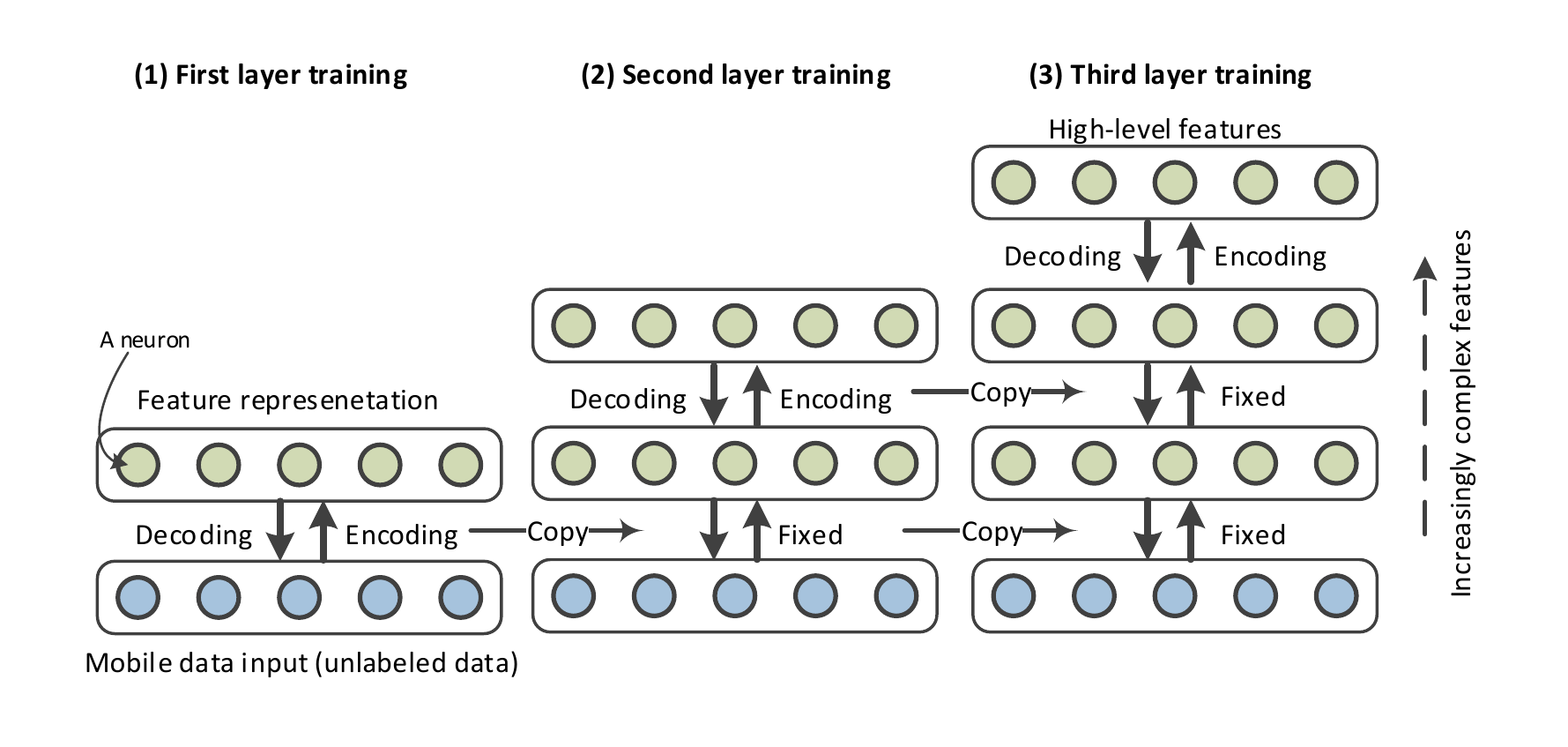}
\par\end{centering}

\caption{Generative layer-wise training of a deep model. Each layer applies nonlinear transformation to its input vector and produces intrinsic
features at its output.\label{fig:greedy_training}}
\end{figure}

Deep learning is a new branch of machine learning which can solve a broad set of complex problem in MBD analytics, e.g., classification and regression. A \emph{deep learning model} consists of simulated neurons and synapses which can be trained to learn hierarchical features from existing MBD samples. The resulting deep model can generalize and process unseen streaming MBD samples.

For simplicity, we present a general discussion of deep learning methods without focusing on the derivations of particular techniques. Nonetheless, we refer interested readers to more technical papers of deep belief networks~\cite{hinton2006fast} and stacked denoising autoencoders~\cite{vincent2010stacked}. A deep model can be scaled to contain many hidden layers and millions of parameters which are difficult to be trained at once. Instead, greedy layer-by-layer learning algorithms~\cite{hinton2006fast,vincent2010stacked} were proposed which basically work as follows:
\begin{enumerate}
\item \emph{Generative layer-wise pre-training}: This stage requires only unlabeled data which is often abundant and cheap to collect in mobile systems using crowdsourcing. Figure~\ref{fig:greedy_training} shows the layer-wise tuning of a deep model. Firstly, one layer of neurons is trained using the unlabeled data samples. To learn the input data structure, each layer includes encoding and decoding functions: The encoding function uses the input data and the layer parameters to generate a set of new features. Then, the decoding function uses the features and the layer parameters to produce a reconstruction of the input data. As a result, a first set of features is generated at the output of the first layer. Then, a second layer of neurons is added at the top of the first layer, where the output of the first layer is fed as input of the second layer. This process is repeated by adding more layers until a suitable deep model is formed. Accordingly, more complex features are learned at each layer based on the features that were generated at its lower layer. 
\item \emph{Discriminative fine-tuning}: The model's parameters which were initialized in the first step are then slightly fine-tuned using the available set of labeled data to solve the problem at hand.
\end{enumerate}

\subsection{Deep Learning Advantages in MBD Analytics}

Deep learning provides solid learning models for MBD analytics. This argument can be supported with the following advantages of using deep learning in MBD analytics:
\begin{itemize}
\item \emph{Deep learning scores high-accuracy results which are a top priority for growing mobile systems}. High-accuracy results of MBD analytics are required for sustainable business and effective decisions. For example, a poor fraud detection results in expensive loss of income for mobile systems. Deep learning models have been reported as state-of-the-art methods in solving many MBD tasks. For example, the authors in~\cite{wang2011deepfi} propose a method for indoor localization using deep learning and channel state information. In~\cite{lane2015can}, deep learning is successfully applied to inference tasks in mobile sensing, e.g., activity and emotion recognition, and speaker identification.
\item \emph{Deep learning generates intrinsic features which are required in MBD analytics}. A \emph{feature} is a measurement attribute extracted from sensory data to capture the underlying phenomenon being observed and enable more effective MBD analytics. Deep learning can automatically learn high-level features from MBD, eliminating the need for handcrafted features in conventional machine learning methods.
\item \emph{Deep Learning can learn from unlabeled mobile data which minimizes the data labeling effort}. In most mobile systems, labeled data is limited, as manual data annotation requires expensive human intervention which is both costly and time consuming. On the other hand, unlabeled data samples are abundant and cheap to collect. Deep learning models utilize unlabeled data samples for generative data exploration during a pre-training stage. This minimizes the need for labeled data during MBD analytics.
\item \emph{Multimodal deep learning}. The ``variety'' aspect of MBD leads to multiple data modalities of multiple sensors (e.g., accelerometer samples, audio, and images). Multimodal deep learning~\cite{ngiam2011multimodal} can learn from multiple modalities and heterogeneous input signals.
\end{itemize}

\subsection{Deep Learning Challenges in MBD Analytics}

Discussing MBD in terms of volume only and beyond the analytical and profit perspectives is incomplete and restricted. Collecting MBD is unprofitable unless suitable learning methods and analytics are utilized in extracting meaningful information and patterns. Deep learning in MBD analytics is slow and can take a few days of processing time, which does not meet the operation requirements of most modern mobile systems. This is due to the following challenges:
\begin{itemize}
\item \emph{Curse of dimensionality}: MBD comes with ``volume'' and ``velocity'' related challenges. Historically, data analytics on small amounts of collected data (a.k.a.~random sampling) was utilized. Despite the low computational burdens of random sampling, it suffers from poor performance on unseen streaming samples. This performance problem is typically avoided by using the full set of available big data samples which significantly increases the computational burdens.
\item \emph{Large-scale deep models}: To fully capture the information on MBD and avoid underfitting, deep learning models should contain millions of free parameters, e.g., a 5-layer deep model with 2000 neurons per layer contains around 20 million parameters. The model free parameters are optimized using gradient-based learning~\cite{hinton2006fast,vincent2010stacked} which is computationally expensive for large-scale deep models.
\item \emph{Time-varying deep models}: In mobile systems, the continuous adaptation of deep models over time is required due to the ``volatility'' characteristic of MBD.
\end{itemize}
To tackle these challenges, we next describe a scalable framework for MBD analytics using deep learning models and Apache Spark.

\section{A Spark-based Deep Learning Framework for MBD Analytics}\label{sec:DL_MBD_spark}

\begin{figure*}
\begin{centering}
\includegraphics[width=0.65\paperwidth,trim=1cm 0.8cm 1cm 0cm]{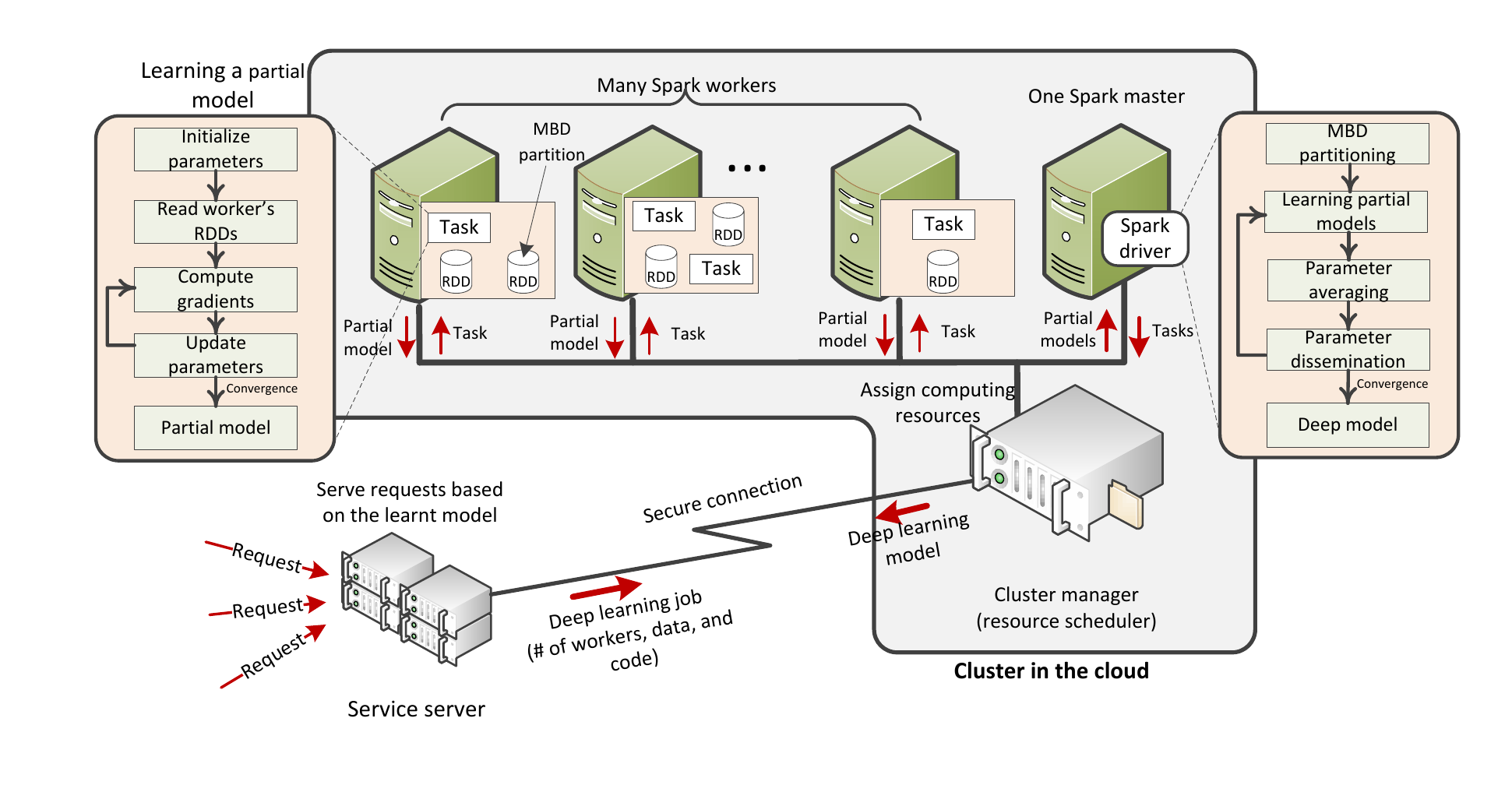}
\par\end{centering}

\caption{A Spark-based framework for distributed deep learning in MBD analytics.\label{fig:proposed_architecture}}
\end{figure*}

Learning deep models in MBD analytics is slow and computationally demanding. Typically, this is due to the large number of parameters of deep models and the large number of MBD samples. Figure~\ref{fig:proposed_architecture} shows the proposed architecture for learning deep models on MBD with Apache Spark. Apache Spark~\cite{spark2015apache} is an open source platform for scalable MapReduce computing on clusters. The main goal of the proposed framework is speeding up MBD decision-making by parallelizing the learning of deep models to a high performance computing cluster. In short, the parallelization of a deep model is performed by slicing the MBD into many partitions. Each partition is contained in a resilient distributed dataset (RDD) which provides an abstraction for data distribution by the Spark engine. Besides data caching, RDDs of a Spark-program natively support fault-tolerant executions and recover the program operations at worker nodes.

In short, our Spark-based framework consists of two main components: (1)~a Spark master and (2)~one or more Spark workers. The master machine initializes an instance of the Spark driver that manages the execution of many partial models at a group of Spark workers. At each iteration of the deep learning algorithm (Figure~\ref{fig:greedy_training}), each worker node learns a partial deep model on a small partition of the MBD and sends the computed parameters back to the master node. Then, the master node reconstructs a master deep model by averaging the computed partial models of all executor nodes.

\subsection{Parallelized Learning Collections}

Learning deep models can be performed in two main steps: (1)~gradient computation, and (2)~parameter update (see~\cite{hinton2006fast,vincent2010stacked} for the mathematical derivation). In the first step, the learning algorithm iterates through all data batches independently to compute gradient updates, i.e., the rate of change, of the model's parameters. In the second step, the model's parameters are updated by averaging the computed gradient updates on all data batches. These two steps fit the learning of deep models in the MapReduce programming model~\cite{dean2012large,zhang2014large}. In particular, the parallel gradient computation is realized as a Map procedure, while the parameter update step reflects the Reduce procedure. The iterative MapReduce computing of deep learning on Apache Spark is performed as follows:
\begin{enumerate}
\item \emph{MBD partitioning}: The overall MBD is first split into many partitions using the \emph{parallelize()} API of Spark. The resulting MBD partitions are stored into RDDs and distributed to the worker nodes. These RDDs are crucial to speedup the learning of deep models as the memory data access latency is significantly shorter than the disk data operations.
\item \emph{Deep learning parallelism}: The solution of a deep learning problem depends on the solutions of smaller instances of the same learning problem with smaller datasets. In particular, the deep learning job is divided into learning stages. Each learning stage contains a set of independent MapReduce iterations where the solution of one iteration is the input for the next iteration. During each MapReduce iteration, a partial model is trained on a separate partition of the available MBD as follows:

\begin{enumerate}
\item \label{enu:map_phase}\emph{Learning partial models}: Each worker node computes the gradient updates of its partitions of the MBD (a.k.a.~the Map procedure). During this step, all Spark workers execute the same Map task in parallel but on different partitions of the MBD. In this way, the expensive gradient computation task of the deep model learning is divided into many parallel sub-tasks.
\item \emph{\label{enu:reduce_phase}Parameter averaging}: Parameters of the partial models are sent to the master machine to build a master deep model by averaging the parameter calculation of all Spark workers (a.k.a.~the Reduce procedure).
\item \emph{Parameter dissemination}: The resulting master model after the Reduce procedure is disseminated to all worker nodes. A new MapReduce iteration is then started based on the updated parameters. This process is continued until the learning convergence criterion is satisfied.
\end{enumerate}
\end{enumerate}
As a result, a well-tuned deep learning model is generated which can be used to infer information and patterns from streaming requests. In the following, we discuss how the proposed framework helps in tackling the key characteristics of MBD.

\subsection{Discussion}

The proposed framework is grounded over deep learning and Apache Spark technologies to perform effective MBD analytics. This integration tackles the challenging characteristics of MBD as follows.
\begin{itemize}
\item \emph{Deep learning}: Deep learning addresses the ``value'' and ``variety'' aspects of MBD. Firstly, deep learning in MBD analytics helps in understanding raw MBD. Therefore, deep learning effectively addresses the ``value'' aspect of MBD. MBD analytics, as discussed in this article, is integral in providing user-customized mobile services. Secondly, deep learning enables the learning from multimodal data distributions~\cite{ngiam2011multimodal}, e.g., concatenated input from accelerometer and light sensors, which is important for the ``variety'' issue of MBD.
\item \emph{Apache Spark}: The main role of the Spark platform in the proposed framework is tackling the ``volume'', ``velocity'', and ``volatility'' aspects of MBD. Essentially, the Spark engine tackles the ``volume'' aspect by parallelizing the learning task into many sub-tasks each performed on a small partition of the overall MBD. Therefore, no single machine is required to process the massive MBD volume as one chunk. Similarly, the Spark engine tackles the ``velocity'' point through its streaming extensions which enables a fast and high-throughput processing of streaming data. Finally, the ``volatility'' aspect is addressed by significantly speeding up the training of deep models. This ensures that the learned model reflects the latest dynamics of the mobile system.
\end{itemize} 
The proposed framework does not directly tackle the ``veracity'' aspect of MBD. This quality aspect requires domain experts to design conditional routines to check the validity of crowdsourced data before being added to a central MBD storage.

\section{Prototyping Context-Aware Activity Recognition System}\label{sec:case_study}

Context-awareness~\cite{perera2014context,lara2013survey} has high impact on understanding MBD by describing the circumstances during which the data was collected, so as to provide personalized mobile experience to end users, e.g.,~targeted advertising, healthcare, and social services. A \emph{context} contains attributes of information to describe the sensed environment such as performed human activities, surrounding objects, and locations. A \emph{context learning model} is a program that defines the rules of mapping between raw sensory data and the corresponding context labels, e.g., mapping accelerometer signals to activity labels. This section describes a proof-of-concept case study in which we consider a context-aware activity recognition system, e.g., detect walking, jogging, climbing stairs, sitting, standing, and lying down activities. We use real-world dataset during the training of deep activity recognition models.

\subsection{Problem Statement}

Accelerometers are sensors which measure proper acceleration of an object due to motion and gravitational force. Modern mobile devices are widely equipped with tiny accelerometer circuits which are produced from electromechanically sensitive elements and generate electrical signal in response to any mechanical motion. The proper acceleration is distinctive from coordinate acceleration in classical mechanics. The latter measures the rate of change of velocity while the former measures acceleration relative to a free fall, i.e., the proper acceleration of an object in a free fall is zero.

Consider a mobile device with an embedded accelerometer sensor that generates proper acceleration samples. Activity recognition is applied to time series data frames which are formulated using a sliding and overlapping window. The number of time-series samples depends on the accelerometer's sampling frequency (in Hertz) and windowing length (in seconds). At time $t$, the activity recognition classifier $f:\mathbf{x}_{t}\rightarrow\mathcal{S}$ matches the framed acceleration data $\mathbf{x}_{t}$ with the most probable activity label from the set of supported activity labels $\mathcal{S}=\left\{ 1,2,\ldots,N\right\} $, where $N$ is the number of supported activities in the activity detection component.

Conventional approaches of recognizing activities require handcrafted features, e.g., statistical features~\cite{lara2013survey}, which are expensive to design, require domain expert knowledge, and generalize poorly to support more activities. To avoid this, a deep activity recognition model learns not only the mapping between raw acceleration data and the corresponding activity label, but also a set of meaningful features which are superior to handcrafted features.

\subsection{Experimental Setup}

\begin{figure*}
\begin{centering}
\subfloat[]{\begin{centering}
\includegraphics[width=0.95\columnwidth,trim=1cm 1.5cm 1cm 0cm]{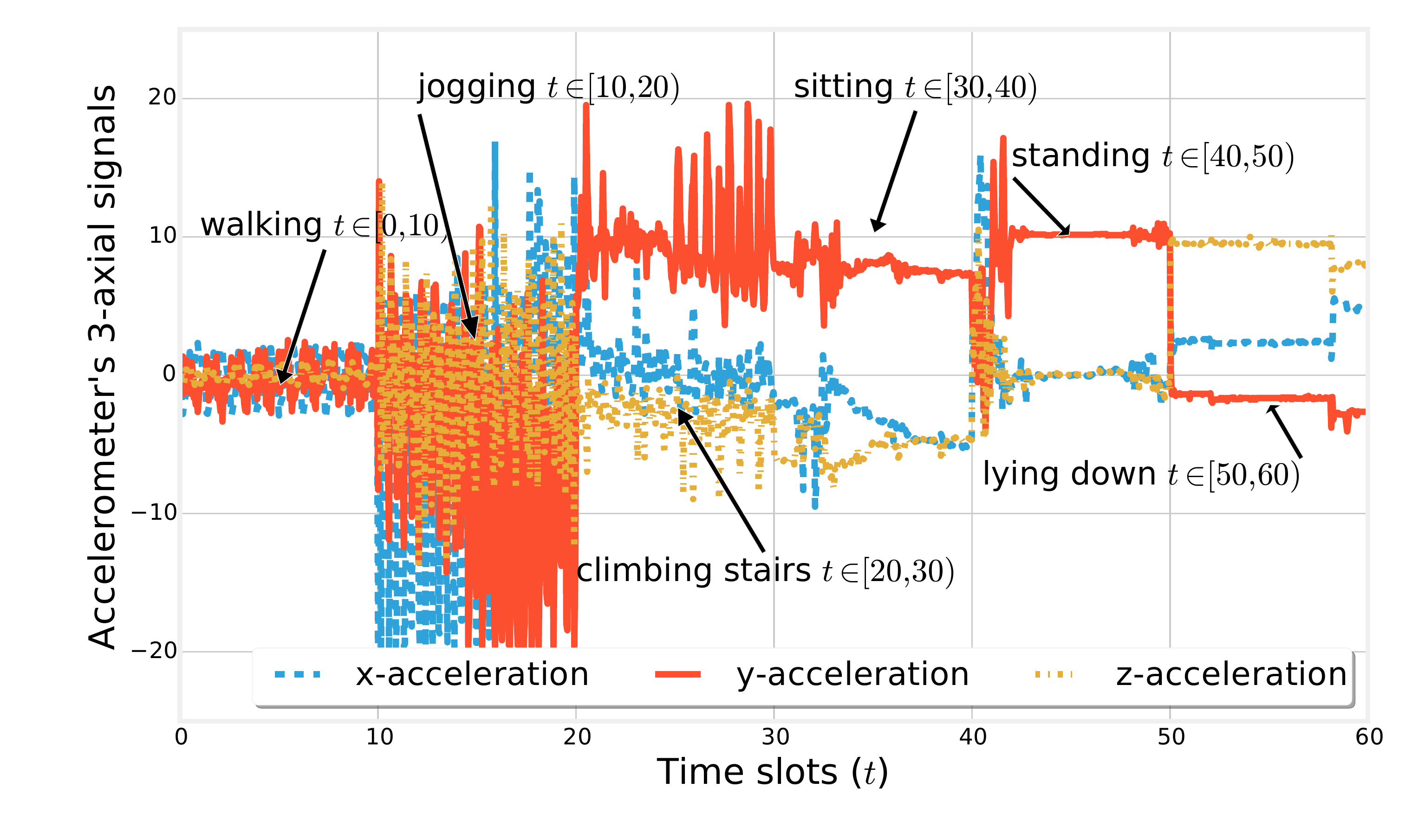}
\par\end{centering}

}
\par\end{centering}

\begin{centering}
\subfloat[]{\begin{centering}
\includegraphics[width=0.85\columnwidth,trim=1cm 1cm 0cm 1cm]{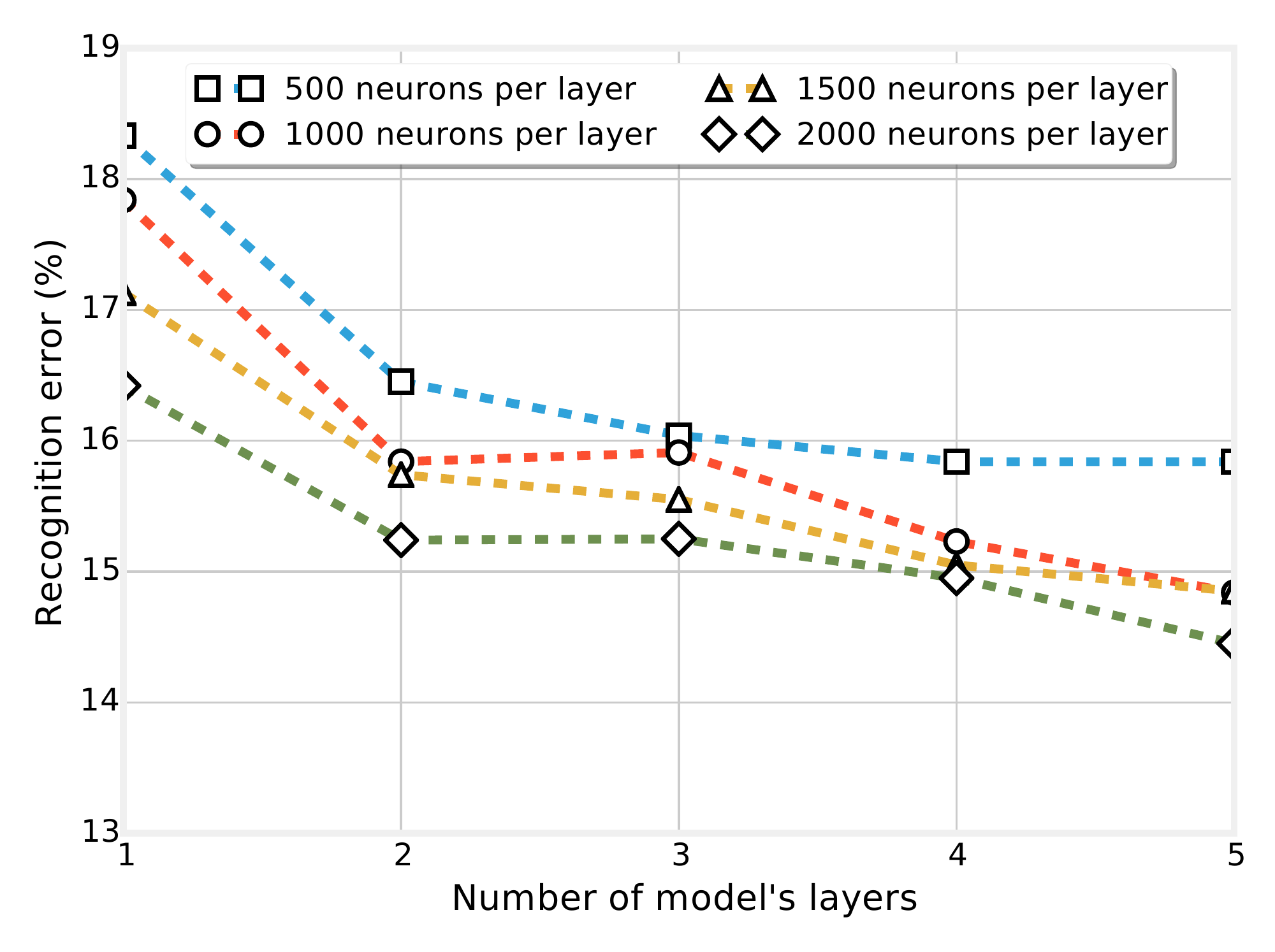}
\par\end{centering}

}\subfloat[]{\begin{centering}
\includegraphics[width=0.85\columnwidth,trim=0cm 1cm 1cm 1cm]{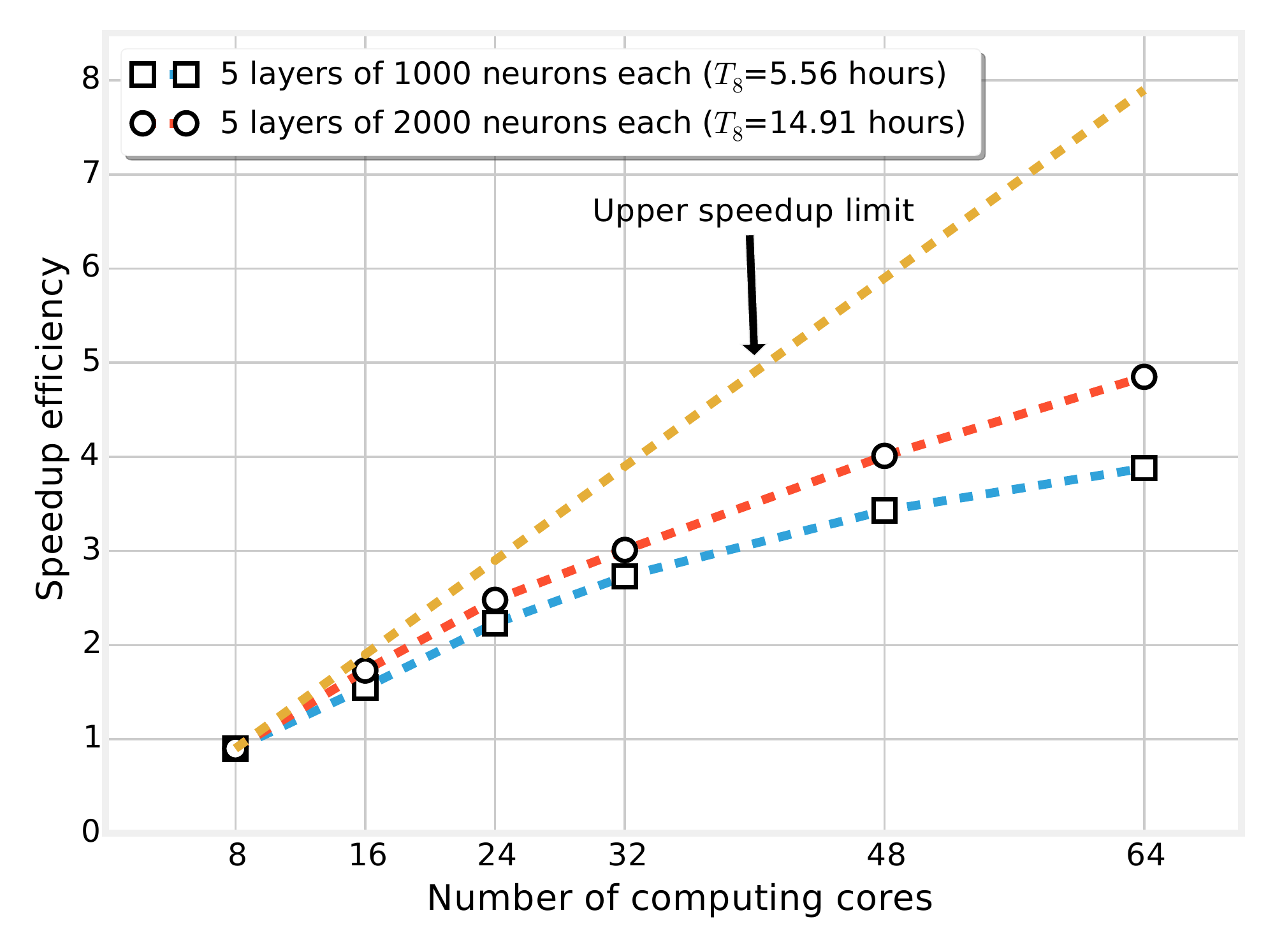}
\par\end{centering}
}
\par\end{centering}

\caption{Experimental analysis. (a)~Accelerometer signal of different human activities. (b)~Recognition accuracy of deep learning models under different deep model setups. (c)~Speedup of learning deep models using the Spark-based framework under different computing cores.\label{fig:analysis_results_err}}
\end{figure*}

In this section, we use the Actitracker dataset~\cite{lockhart2011design} which includes accelerometer samples of $6$~conventional activities (walking, jogging, climbing stairs, sitting, standing, and lying down) from $563$ crowdsourcing users. Figure~\ref{fig:analysis_results_err}~(a) plots accelerometer signals of the 6 different activities. Clearly, high frequency signals are sampled for activities with active body motion, e.g., walking, jogging, and climbing stairs. On the other hand, low frequency signals are collected during semi-static body motions, e.g., standing, sitting, and lying down. The data is collected using mobile phones with $20$Hz of sampling rate, and it contains both labeled and unlabeled data of $2,980,765$ and $38,209,772$~samples, respectively. This is a real-world example of the limited number of labeled data compared with unlabeled data as data labeling requires manual human intervention. The data is framed using a $10$-sec windowing function which generates $200$~samples of time-series samples. We first pre-train deep models on the unlabeled data samples only, and we then fine-tune the models on the labeled dataset. To enhance the activity recognition performance, we use the spectrogram of the acceleration signal as input of the deep models. Basically, different activities contain different frequency contents which reflect the body dynamics and movements.

We implemented the proposed framework on a shared cluster system (https://www.acrc.a-star.edu.sg) running the load sharing facility (LSF) management platform and RedHat Linux. Each node has $8$~cores (Intel Xeon $5570$~CPU with clock speed of $2.93$Ghz) and a total of $24$GB RAM. In our experiments, we set the cores in multiples of $8$ to allocate the entire node's resources. One partial model learning task is initialized per each computing core. Each task learns using a data batch consisting of 100~samples for 100~iterations. Clearly, increasing the number of cores results in quicker training of deep models. Finally, it important to note that distributed deep learning is a strong type of regularization. Thus, regularization terms, such as the sparsity and dropout constraints, are not recommended to avoid the problem of underfitting.

\subsection{Experimental Results}

\subsubsection{The impact of deep models}

Figure~\ref{fig:analysis_results_err}~(b) shows the activity recognition error under different setups of deep models (number of hidden layers and number of neurons at each layer). Specifically, the capacity of a deep model to capture MBD structures is increased when using deeper models with more layers and neurons. Nonetheless, using deeper models evolves a significant increase in the learning algorithm's computational burdens and time. An accuracy comparison of deep activity recognition models and other conventional methods is shown in Table~\ref{tab:comparison_conventional_methods}. In short, these results clarify that (1)~deep models are superior to existing shallow context learning models, and (2)~the learned hierarchical features of deep models eliminate the need for handcrafted statistical features in conventional methods. In our implementation, we use early stopping to track the model capacity during training, select the best parameters of deep models, and avoid overfitting. The underfitting is typically avoided by using deeper models and more neurons per layer, e.g., 5 layers with 2000 neurons per layer. Next, a speedup analysis is presented to show the importance of the Spark-based framework for learning deep models on MBD.

\subsubsection{The impact of computing cores}

The main performance metric of cluster-based computing is the task speedup metric. In particular, we compute the speedup efficiency as $\frac{T_{8}}{T_{M}}$, where $T_{8}$ is the computing time of one machine with $8$ cores, and $T_{M}$ is the computing time under different computing power. Figure~\ref{fig:analysis_results_err}~(c) shows the speedup in learning deep models when the number of computing cores is varied. As the number of cores increases, the learning time decreases. For example, learning a deep model of 5 layers with 2000 neurons per layer can be trained in $3.63$ hours with 6 Spark workers. This results in the speedup efficiency of $4.1$ as compared to a single machine computing which takes $14.91$ hours.

\subsubsection{MBD veracity}

\begin{figure}
\begin{centering}
\includegraphics[width=0.85\columnwidth,trim=1cm 0.8cm 1cm 0cm]{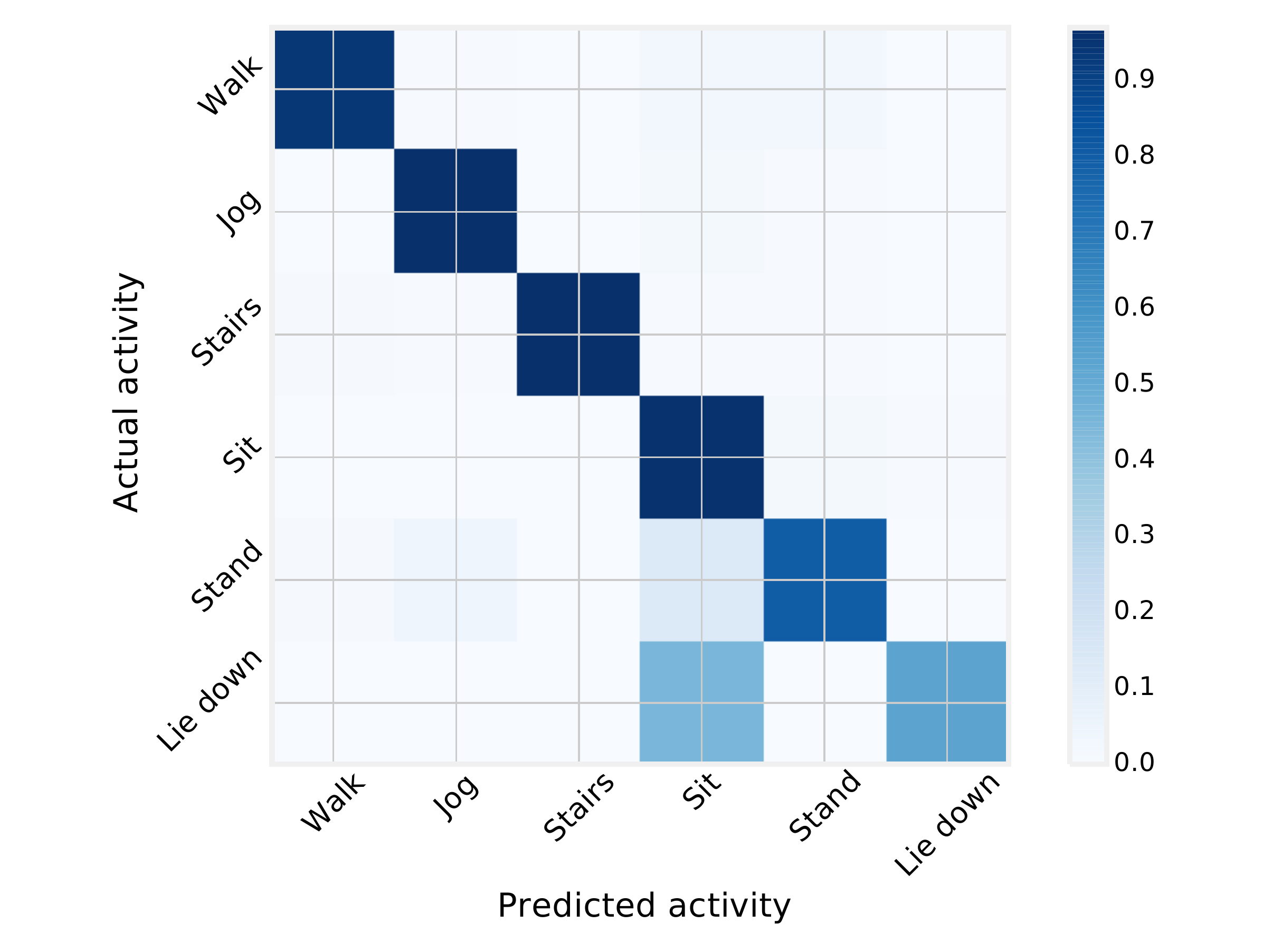}
\par\end{centering}
\caption{Normalized confusion matrix of a deep model (5 layers of 2000 neurons each).\label{fig:confession_matrix}}
\end{figure}

A normalized confusion matrix of a deep model is shown in Figure~\ref{fig:confession_matrix}. This confusion matrix shows the high performance of deep models on a per-activity basis (high scores at the diagonal entries). The incorrect detection of the ``sitting'' activity instead of the ``lying down'' activity is typically due to the different procedures in performing the activities by crowdsourcing users. This gives a real-world example of the ``veracity'' characteristic of MBD, i.e.,~uncertainties in MBD collection.

\begin{table}
\caption{Activity recognition error of deep learning and other conventional methods used in~\cite{weiss2012impact}. The conventional methods use handcrafted statistical features.\label{tab:comparison_conventional_methods}}

\centering{}%
\begin{tabular}{|>{\centering}p{4cm}|c|}
\hline 
\textbf{\noun{Method}} & \textbf{\noun{Recognition error (\%)}}\tabularnewline
\hline 
\hline 
Multilayer perceptrons & 32.2\tabularnewline
\hline 
Instance-based learning & 31.6\tabularnewline
\hline 
Random forests & 24.1\tabularnewline
\hline 
\textbf{Deep models (5 layers of 2000 neurons each)} & \textbf{14.4}\tabularnewline
\hline 
\end{tabular}
\end{table}

In the next section, we identify some notable future research directions in MBD collection, labeling, and economics.

\section{Future Work}\label{sec:future_work}

Based on the proposed framework, the following future work can be further pursued.

\subsection{Crowd Labeling of MBD}

A major challenge facing MBD analysts is the limited amounts of labeled data samples as data labeling is typically a manual process. An important research direction is proposing crowd labeling methods for MBD. The crowd labeling can be designed under two main schemes: (1)~paid crowd labeling, and (2)~embedded crowd labeling. In the paid crowd labeling, the crowdsourcing mobile users annotate mobile data and are accordingly paid based on their labeling performance and speed. Under this paid scheme, optimal budget allocation methods are required. In the embedded crowd labeling, data labeling can be also achieved by adding labeling tasks within mobile application functional routines, e.g., CAPTCHA-based image labeling~\cite{von2008recaptcha}. Here, the mobile users can access more functions of a mobile application by indirectly helping in the data labeling process. More work is required for designing innovative methods for embedded crowd labeling without disturbing the user experience or harming the mobile application's main functionality.

\subsection{Economics of MBD}

MBD, as discussed in this article, is about extracting meaningful information and patterns from raw mobile data. This information is used during decision making and to enhance existing mobile services. An important research direction is proposing business models, e.g., pricing and auction design~\cite{klemperer2004auctions}, for selling and buying MBD among mobile organizations and parties.

\subsection{Privacy and MBD Collection}

As MBD is people-centric, mobile users would be concerned about the risks of sharing their personal mobile data with a service server. Thus, a low percentage of users will opt out of sharing their personal data unless trustworthy privacy mechanisms are applied. Meanwhile, anonymized data collection, i.e., data that could not be used to identify individuals, is adopted by many services. An alternative research direction is proposing fair data exchange models which encourage the sharing of mobile data in return of rewarding points, e.g., premium membership points.

\section{Conclusions}\label{sec:conclusions}

In this article, we have presented and discussed a scalable Spark-based framework for deep learning in mobile big data analytics. The framework enables the tuning of deep models with many hidden layers and millions of parameters on a cloud cluster. Typically, deep learning provides a promising learning tool for adding value by learning intrinsic features from raw mobile big data. The framework has been validated using a large-scale activity recognition system as a case study. Finally, important research directions on mobile big data have been outlined.

\section*{Acknowledgment}

This work was supported by the A{*}STAR Computational Resource Centre through the use of its high performance computing facilities. We thank Ahmed Selim, Trinity College Dublin, for valuable discussions in the early stages of the study.

\bibliographystyle{IEEEtran}
\bibliography{mobile_big_data}

\end{document}